\documentclass[aps,prl,twocolumn,showpacs,superscriptaddress]{revtex4-1}
\usepackage{graphicx,amsmath,amssymb,amsfonts}
\usepackage[latin1]{inputenc}
\usepackage{array}
\usepackage{multirow}
\usepackage{float}
\usepackage{color}
\usepackage[normalem]{ulem}
\usepackage{booktabs}

\begin{document}

\title{Multiple magnetic transitions and complex magnetic structures in Fe$_2$SiSe$_4$ with the sawtooth lattice}

\author{Feihao Pan}
\thanks{These authors contributed equally to this work}
\affiliation{Laboratory for Neutron Scattering and Beijing Key Laboratory of Optoelectronic Functional Materials and MicroNano Devices, Department of Physics, Renmin University of China, Beijing 100872, China}
\author{Xunwu Hu}
\thanks{These authors contributed equally to this work}
\affiliation{State Key Laboratory of Optoelectronic Materials and Technologies, Guangdong Provincial Key Laboratory of Magnetoelectric Physics and Devices, Center for Neutron Science and Technology, School of Physics, Sun Yat-sen University, Guangzhou, 510275, China}

\author{Jiale Huang}
\author{Bingxian Shi}
\author{Jinchen Wang}
\author{Juanjuan Liu}
\author{Hongxia Zhang}
\author{Daye Xu}
\affiliation{Laboratory for Neutron Scattering and Beijing Key Laboratory of Optoelectronic Functional Materials and MicroNano Devices, Department of Physics, Renmin University of China, Beijing 100872, China}

\author{Hongliang Wang}
\author{Lijie Hao}
\affiliation{ China Institute of Atomic Energy, PO Box-275-30, Beijing 102413, China}

\author{Peng Cheng}
\email[Corresponding author: ]{pcheng@ruc.edu.cn}
\affiliation{Laboratory for Neutron Scattering and Beijing Key Laboratory of Optoelectronic Functional Materials and MicroNano Devices, Department of Physics, Renmin University of China, Beijing 100872, China}
\author{Dao-Xin Yao}
\email[Corresponding author: ]{yaodaox@mail.sysu.edu.cn}
\affiliation{State Key Laboratory of Optoelectronic Materials and Technologies, Guangdong Provincial Key Laboratory of Magnetoelectric Physics and Devices, Center for Neutron Science and Technology, School of Physics, Sun Yat-sen University, Guangzhou, 510275, China}

\begin{abstract}
The sawtooth lattice shares some structural similarities with the kagome lattice and may attract renewed research interest. Here, we report a comprehensive study on the physical properties of Fe$_2$SiSe$_4$, an unexplored member in the olivine chalcogenides with the sawtooth lattice of Fe. Our results show that Fe$_2$SiSe$_4$ is a magnetic semiconductor with band gap of 0.66~eV. It first undergoes an antiferromagnetic transition at T$_{m1}$=110~K, then an ferrimagnetic-like one at T$_{m2}$=50~K and finally a magnetic transition at T$_{m3}$=25~K which is likely driven by the thermal populations of spin-orbit manifold on the Fe site. Neutron diffraction analysis reveals a non-collinear antiferromagnetic structure with propagation vector $\mathbf{q_1}$=(0,0,0) at T$_{m2}$$\textless$T$\textless$T$_{m1}$. Interestingly, below T$_{m2}$, an additional antiferromagnetic structure with $\mathbf{q_2}$=(0,0.5,0) appears and Fe$_2$SiSe$_4$ exhibits a complex double-$\mathbf{q}$ magnetic structure which has never been observed in sawtooth olivines. Density functional theory calculations suggest this complex noncollinear magnetic structure may originate from the competing antiferromagnetic interactions for both intra- and inter-chain in the sawtooth lattice. Furthermore, band structural calculations show that Fe$_2$SiSe$_4$ has quasi-flat band features near the valence and conduction bands. Based on the above results, we propose Fe$_2$SiSe$_4$ as a new material platform to condensed matter researches. 
\end{abstract}

\maketitle




\section{Introduction}
Materials with the sawtooth lattice have attracted attentions from the condensed matter physics community for several reasons. Firstly, the sawtooth antiferromagnetic (AFM) chain with corner-sharing triangles of spins represents one of the
fundamental models of geometrically frustrated quantum magnetism as that in triangular and kagome lattices\cite{Cava,Cava2}. Secondly, the sawtooth lattice also exhibits flat-band feature\cite{Flatband2010} which may give rise to high thermoelectric performance\cite{Flatband2015}, quantum topological phase\cite{topo1,topo2,topo3}, flat-band spin dynamics and phonon anomalies\cite{flatspin}. Besides, a two-dimensional kagome lattice can be viewed as the combination of one-dimensional sawtooth chains. Therefore the recent discoveries of novel physical properties in kagome material\cite{Fe3Sn2,CoSnS,CsVSb,KagomeSC,Yin_2021} may generate renewed interest in sawtooth material since these two kinds share structural similarities and connections.

The A$_2$BX$_4$ (A=Mn, Fe, Ni; B=Si, Ge; X=O,S,Se,Te) olivines represent a large material family where the transitional-metal atoms form a sawtooth lattice. Most members in this series are ferrimagnets or antiferromagnets whose spin structure could be described by magnetic propagation vector $\mathbf{q}$=(0,0,0)\cite{Fe2SiO4_1993,Fe2SiS4}. Among them, magnetic frustration is observed for Mn$_2$SiSe$_4$\cite{Mn2SiSe4}. Mn$_2$SiS$_4$ and Mn$_2$GeS$_4$ are reported to exhibit anomalous magnetic properties result from the quantum fluctuations near a spin-flop bicritical point\cite{QCP214}. Mn$_2$GeO$_4$ is identified as a functional material that exhibit coupled ferromagnetism and ferroelectricity\cite{Mn2GeO4}. Fe$_2$GeS$_4$ and Fe$_2$GeSe$_4$ are theoretically proposed as promising candidates with good thermoelectric performance due to the quasi-flat-band feature\cite{Flatband2015}. These findings highlight the exotic physical properties which are closely related to the peculiar sawtooth lattice.

In A$_2$BX$_4$ series, we noticed that although the magnetic properties of Fe$_2$SiO$_4$ and Fe$_2$SiS$_4$ have been reported\cite{Fe2SiO4_2007,Fe2SiS4}, Fe$_2$SiSe$_4$ remains an unexplored member whose chemical phase is even missing in the current inorganic crystal structure database (ICSD). So it would be interesting to check whether the Fe$_2$SiSe$_4$ phase with sawtooth lattice really exists. In this paper, we report the synthesize of Fe$_2$SiSe$_4$ single crystals which is characterized by sawtooth lattice of Fe. Using magnetization, heat capacity, neutron scattering techniques and band structure calculations, Fe$_2$SiSe$_4$ is identified as a magnetic semiconductor with multiple magnetic transitions, non-collinear double-$\mathbf{q}$ magnetic structures and quasi-flat-band. The underlying physical mechanism for the magnetic properties and potential applications of Fe$_2$SiSe$_4$ are discussed combined with the results of density functional theory (DFT) calculations.

\section{Experimental methods}

Single crystals of Fe$_2$SiSe$_4$ were grown by chemical vapor transport method. The pure powder of Fe, Si and Se were mixed in molar ratio 1:2:4 (total mass 1~g), put into a quartz tube (inner diameter of 12~mm, length 10~cm) with 100~mg iodine. Then the quartz tube was evacuated to 3.0$\times$10$^{-3}$Pa and sealed before put into a two-zone tube furnace. The quartz tube was heated to 590$\,^{\circ}\mathrm{C}$ in the raw material end and 660$\,^{\circ}\mathrm{C}$ in the other end in 750 minutes, then maintained at the temperatures for 5760 minutes. The next step is a so-called "temperature reversing process" in which the two ends switch temperatures in 70 minutes and held in the new temperatures for 12.5 days before cooled with the furnace. The shining black single crystals of Fe$_2$SiSe$_4$ appeared in the final cold end with typical dimensions of 1~mm$\times$2~mm$\times$0.5~mm.

X-ray diffraction (XRD) patterns of powder samples were collected from a Bruker D8 Advance X-ray diffractometer using Cu
K$_{\alpha}$ radiation. Magnetization measurements were carried
out in Quantum Design MPMS3. Resistivity and heat capacity of the samples were measured on Quantum Design PPMS-14T. The powder neutron diffraction experiments were carried out on Xingzhi cold neutron triple-axis spectrometer at the China Advanced Research Reactor (CARR). For neutron experiments on Xingzhi, the incident neutron energy was fixed at $16$~meV with a neutron velocity selector used upstream to remove higher order neutrons\cite{XingZhi}. About 3~g Fe$_2$SiSe$_4$ powders (crushed from single crystals) were used in neutron experiments. The program FullProf Suite package was used in the Rietveld refinement of neutron powder diffraction
data\cite{RODRIGUEZCARVAJAL199355,Rietveld.a07067}. 

The electronic structure and magnetic properties calculations were performed using the DFT as implemented in the Vienna {\it ab initio} simulation package (VASP) code\cite{prb47, prb54}. The generalized gradient approximation (GGA) in the form of Perdew-Burke-Ernzerhof (PBE)\cite{prl77} is used for exchange-correlation functional. The projector augmented-wave (PAW) method\cite{prb50} with a 300 eV plane-wave cutoff energy is employed. The valence electrons configurations for each atom are $3d^{7}4s^{1}$ for Fe, $3s^{2}3p^{2}$ for Si, and $3d^{10}4s^{2}4p^{4}$ for Se. A $\Gamma $-centered 3 $\times $ 2 $\times $ 5 k-points mesh within the Monkhorst-Pack scheme is used for Brillouin zone sampling. The Hubbard {\it U} \cite{prb57} for the 3d electrons of Fe is chosen as 2 eV to reproduce the experimental magnetic moments and bandgap of Fe$_2$SiSe$_4$. A 1 $\times $ 2 $\times $ 1 supercell is adopted due to the non-collinear AFM structure. The direction of the magnetic moment is constrained, that is, the superimposed double-$\mathbf{q}$ magnetic structure is adopted. All calculations are performed using the experimental structural parameters. Convergence criteria employed for both the electronic self-consistent iteration and the ionic relaxation are set to $10^{-6}$ eV and 0.01 eV/$\mathring{A}$, respectively. Heisenberg exchange interactions are calculated by the four-states method\cite{prb84}. A 1 $\times $ 2 $\times $ 2 supercell is adopted to avoid spurious interaction by the periodic boundary conditions.

\begin{figure}
	\includegraphics[width=8cm]{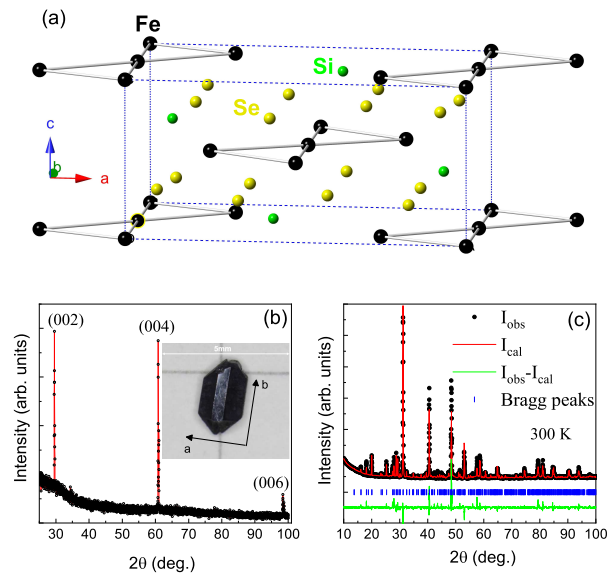}
	\caption {(a) Crystal structure of Fe$_2$SiSe$_4$. (b)  The x-ray reflections from the $ab$-plane of Fe$_2$SiSe$_4$ single crystal. The inset shows the photo of one crystal. (c) Room temperature XRD patterns of powders (crushed from single crystals) and the Rietveld refinement results.} \label{Fig1}
\end{figure}

\section{Results and discussions}

\subsection{Crystal structure and magnetization}

In Fig. 1(c), the XRD patterns on powder samples (crushed from single crystals) and Rietveld refinement confirm Fe$_2$SiSe$_4$ adopts the orthorhombic symmetry with space group $Pnma$ (No.62), which belongs to the olivine-type structure, same as Fe$_2$SiS$_4$ and Fe$_2$SiO$_4$\cite{Fe2SiO4_1993,Fe2SiS4}. The obtained lattice parameters are $a=13.032$\text{\AA}, $b=7.549$\text{\AA} and $c=6.123$\text{\AA}. The x-ray reflection pattern from the $ab$-plane of Fe$_2$SiSe$_4$ single crystal in Fig. 1(b) also confirms the above results. As shown in Fig. 1(a), The Fe atoms form infinite sawtooth chains along the $b$-axis.  

The Laue diffraction patterns allow us to distinguish the major crystal axes with respect to the three-dimensional shape of Fe$_2$SiSe$_4$ single crystal. So the temperature dependent magnetization along $a$-, $b$- and $c$-axis were performed respectively, the results are shown in Fig. 2. Three successive magnetic transitions could be identified. At T$_{m1}$=110~K, a cusp appears which is most prominent under $H\parallel b$ indicating an AFM transition. At lower temperatures, a second magnetic transition occurs at T$_{m2}$=50~K. Its temperature dependent magnetization behavior seems to be AFM-like under $H\parallel b$ while ferromagnetic-like along other field directions. Then at T$_{m3}$=25~K, the magnetization anomalies along all three directions suggest the existence of a third magnetic transition.

\begin{figure}
	\includegraphics[width=8cm]{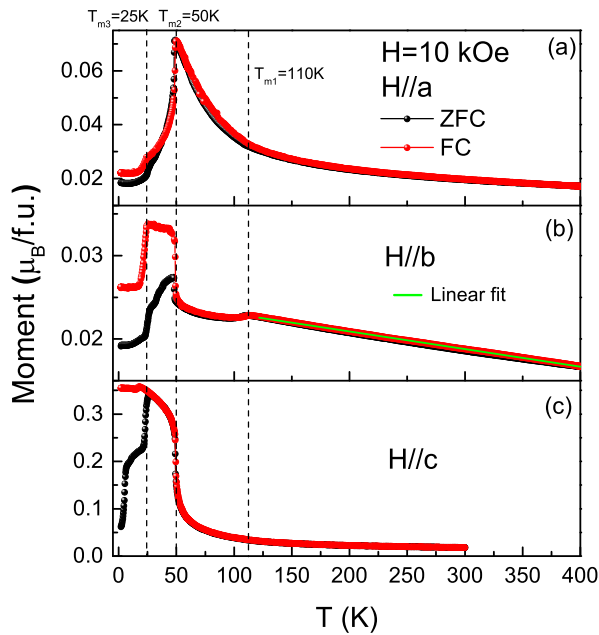}
	\caption {Temperature dependent magnetizations with applied field along the $a$, $b$ and $c$-axis are shown respectively in (a), (b) and (c).} \label{Fig2}
\end{figure}

The temperature dependent magnetization shows strong anisotropic behavior along three different directions. Even in the paramagnetic region, the magnetization under $H\parallel a$ or $H\parallel c$ exhibits typical Curie-Weiss (CW) behavior while that under $H\parallel b$ shows an anomalous linear temperature dependence up to 400~K. The CW fit of the high-temperature magnetization data yields $\mu_{eff}/Fe=4.02~\mu_{B}$ and $\theta_{CW}$=-66~K for $H\parallel a$, $\mu_{eff}/Fe=4.08~\mu_{B}$ and $\theta_{CW}$=-51~K for $H\parallel c$. This result indicates dominant antiferromagnetic correlations in Fe$_2$SiSe$_4$. On the other hand, the T-linear magnetization under $H\parallel b$ is abnormal. Similar behavior has been observed in Fe-based superconductors and parent compound\cite{Tlinear1,Tlinear2,TLinear3}, which is explained as a result of AFM spin fluctuations in the normal state\cite{Zhang_2009}. Further theoretical works are needed to check if this explanation applies to Fe$_2$SiSe$_4$.

\subsection{Magnetic structure}

Since the T-dependent magnetization indicates the existence of three possible magnetic transitions, powder neutron diffraction on Fe$_2$SiSe$_4$ was performed at 150~K, 80~K, 35~K and 3.5~K to determine the magnetic structure. The emergence of new peaks and large enhancement of the intensities for some nuclear peaks could be observed in the neutron diffraction spectra at lower temperature comparing with that at 150~K (Fig. S1). This allows us to identify plenty of magnetic Bragg peaks and determine the magnetic structures at different temperatures through Rietveld refinement. The detailed neutron diffraction data, representational analysis and refinement process are presented in Supplemental Material\cite{Supple}. The main results are shown in Fig. 3.

\begin{figure}
	\includegraphics[width=8cm]{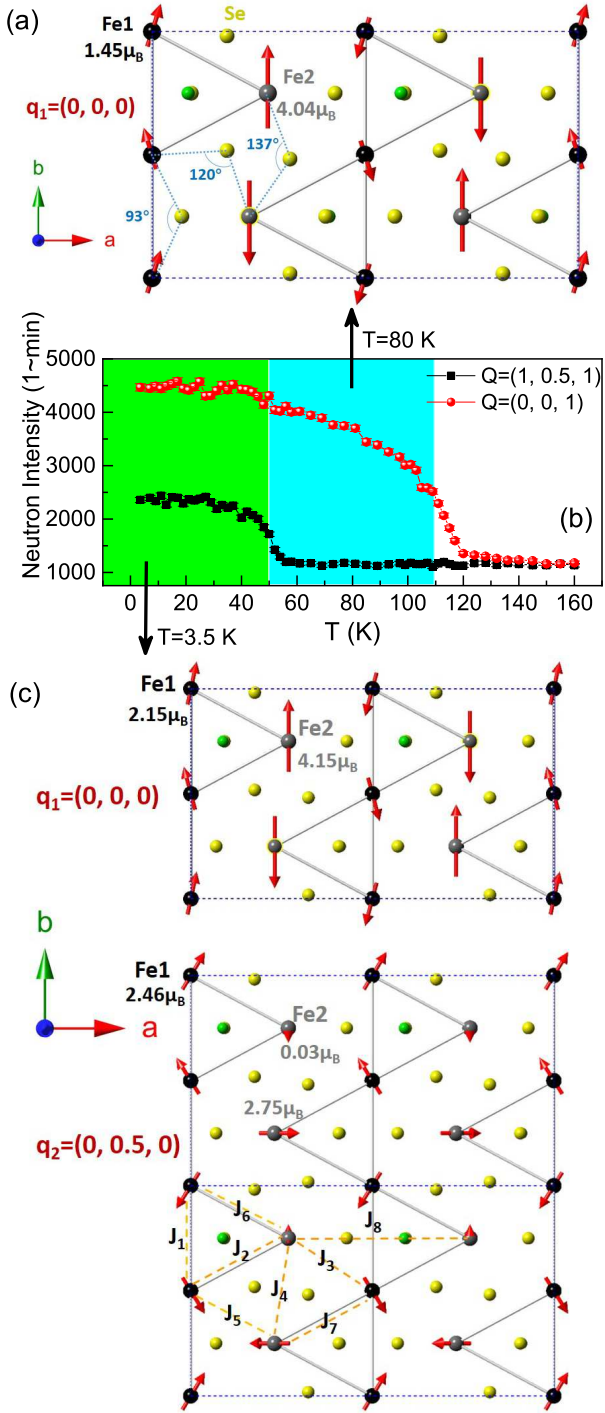}
	\caption {(a) Schematics of refined magnetic structure of Fe$_2$SiSe$_4$ at 80~K. (b) The temperature-dependent intensities of magnetic peaks (0,0,1) and (1,0.5,1). (c) The refined magnetic structures of Fe$_2$SiSe$_4$ at 3.5~K with different propagation vectors are illustrated respectively.} \label{fig3}
\end{figure}

First of all, as shown in Fig. 4(b), the intensities of Bragg peak (0,0,1) could be served as the magnetic order parameter for the first AFM transition at T$_{m1}$. Similarly, the temperature dependent intensity of (1,0.5,1) signifies new magnetic order develops below T$_{m2}$. It should be noted there is almost no difference for the diffraction patterns at 35~K and 3.5~K within our instrumental resolution. Therefore the magnetic transition T$_{m3}$ may not change the zero-field magnetic structure. 

Secondly, at 80~K, the non-collinear AFM structure with magnetic wave-vector $\mathbf{q_1}$=(0,0,0) is identified and illustrated in Fig. 3(a). The Fe ions occupy two inequivalent Wyckoff positions. We label the ones at $4a$ site as Fe1 (black spheres) and that at $4c$ sites as Fe2 (gray spheres). The magnetic moments all lie in the $ab$-plane and the data refinement yields a total moment of 1.45$\mu_{B}$/Fe for Fe1 and 4.04$\mu_{B}$/Fe for Fe2. The latter points parallel to the $b$-axis while the former is canted to the $a$-axis to a certain degree. This magnetic structure is quite similar as that in olivine-type Fe$_2$SiO$_4$ and Fe$_2$SiS$_4$\cite{Fe2SiO4_1993,Fe2SiS4}. Therefore similar physical interpretation of this magnetic structure considering the indirect superexchange interactions between Fe cations via Se atoms could be given. As marked in Fig. 3(a), the Fe2-Se-Fe2 angle is more closer to 180$^{\circ}$ and gives rise to a strong AFM interaction, while the angle of Fe1-Se-Fe1 is much smaller which may largely reduce the AFM interaction\cite{Fe2SiS4}. A competition between crystal-field anisotropy and AFM exchange via spin-orbit coupling results in the canting of Fe1 moments\cite{Fe2SiO4_1993}. 

Thirdly, at 3.5~K, the above magnetic model with $\mathbf{q_1}$=(0,0,0) can only partially fit the neutron diffraction patterns and could not explain new magnetic peaks appear below T$_{m2}$ (Fig. S1). The most prominent one is indexed as (1,0.5,1) whose temperature dependent intensities are shown in Fig. 3(b). The new magnetic peaks developed below T$_{m2}$ can be well defined by a new propagation vector $\mathbf{q_2}$=(0,0.5,0). We found that only a combined model which includes both the magnetic structure with $\mathbf{q_1}$ and that with $\mathbf{q_2}$ could achieve a good fit to the data collected at 3.5~K. The refined two magnetic structures with different wave-vector are illustrated in Fig. 3(c). One can see that the one with $\mathbf{q_1}$ is quite similar as that at 80~K, except for slightly increased ordered moment and different canting angle for Fe1 moment. The magnetic structure with $\mathbf{q_2}$ is also non-collinear and more complex, the magnetic unit cell is doubled along $b$-axis. The moments of Fe2 have two different values and one of them is almost negligible (0.03~$\mu_{B}$).

The magnetic order at 3.5~K have two different propagation wave-vectors. These two modulations may either reside in different domains independently, known as the multi-domain state, or coexist in a single domain in the form of
a superimposed double-$\mathbf{q}$ state. These two states may be indistinguishable in diffraction experiments performed on powder samples. The multi-$\mathbf{q}$ magnetic order is not a common case in magnetic material. Some famous examples include double-$\mathbf{q}$ spin-density wave in Fe-based superconductor Sr$_{1-x}$Na$_x$Fe$_2$As$_2$ with tetragonal crystal symmetry\cite{DoubleQ} and triple-$\mathbf{q}$ magnetic state in Na$_2$Co$_2$TeO$_6$ with hexgonal crystal symmetry\cite{Tripleq}. For Fe$_2$SiSe$_4$, whether there is a double-domain state or a double-$\mathbf{q}$ state, this state in a orthorhombic crystal symmetry is particularly rarely observed.

To explore the underlying mechanism for the complex magnetic structure of Fe$_2$SiSe$_4$ at 3.5~K, we investigate its exchange couplings based on a simple Heisenberg model:
\begin{equation}
	H= \sum_{i<j}J_{ij}S_i S_j
\end{equation}
Our DFT calculations reveal the values of exchange coupling constant $J_i$ between Fe ions at different sites (illustrated in Fig. 3(c)) and the results are listed in Table I. 
We can see that all of the $J_i$ values are positive, which means each side of the sawtooth chains are AFM coupling. The intra-chain exchange coupling $J_1$, $J_2$, $J_6$, and $J_7$ are connected by an exchange path Fe-Si-Se-Fe with the distance {\it d}($J_1$) = 3.78 $\mathring{A}$. The inter-chain exchange coupling $J_3$, $J_4$, $J_5$ are instead realized through exchange path Fe-Se-Se-Fe with the distance {\it d}($J_4$) = 4.89 $\mathring{A}$. Furthermore, the exchange coupling values of inter-chain are almost twice as large as those of intra-chain. This is probably due to the strong hybridization of Se 4$p$ and Fe 3$d$ orbits near the Fermi level. A very large Fe-Fe distance {\it d}($J_8$) = 7.17 $\mathring{A}$ results in a much weaker magnetic interaction. Therefore, the competition of AFM interactions on different sides of sawtooth chain combined with the inter-chain exchange interaction may be responsible for a complex noncollinear double-$\mathbf{q}$ magnetic order in Fe$_2$SiSe$_4$. Based on this model, we employ a superimposed double-$\mathbf{q}$ magnetic structure to constrain the direction of the magnetic moment. The calculated spin configuration at low temperature of Fe$_2$SiSe$_4$ is shown in Table S9 and reach qualitative consistent with the superimposed double-$\mathbf{q}$ magnetic structure determined experimentally(Table S8).
\begin{table}[htbp]
	\caption{Calculated magnetic exchange coupling constants between different Fe sites, positive value means AFM coupling.} \label{tab.1}
	\begin{center}
		\begin{tabular}{ccccccccc}
			Constant & $J_1$ & $J_2$ & $J_3$ & $J_4$ & $J_5$ & $J_6$ & $J_7$ & $J_8$ \\
			\hline
			Value (meV) & 8 & 9 & 15 & 12 & 15 & 9 & 8 &0.3 \\
					\end{tabular}
	\end{center}
\end{table}

\begin{figure*}[htbp]
	\centering
	\includegraphics[width=15cm]{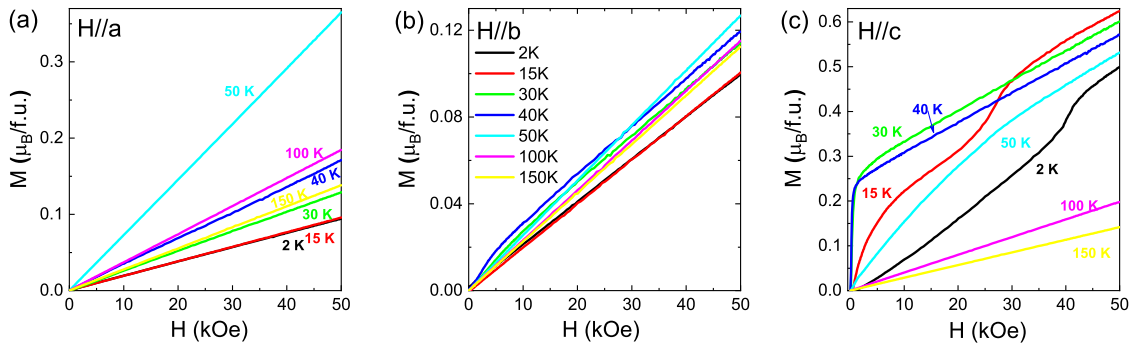}
	\caption {Isothermal magnetizations under magnetic field applied along $a$- ,$b$- and $c$-axis at selected temperatures are shown respectively.} \label{fig4}
\end{figure*}

\begin{figure}
	\includegraphics[width=7cm]{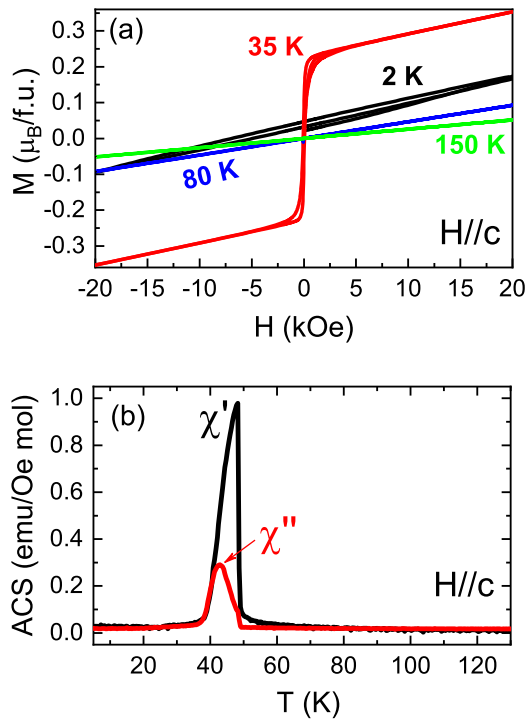}
	\caption {(a) Magnetic hysteresis under $H\parallel c$ for Fe$_2$SiSe$_4$ at selected temperatures. (b) The temperature dependent ac susceptibilities measured under an oscillated AC field of 5.0 Oe applied along $c$-axis.} \label{Fig4}
\end{figure}

\subsection{Isothermal magnetization, AC susceptibility and heat capacity}

Further measurements on Fe$_2$SiSe$_4$ single crystal were performed to get insights on its magnetic state. As shown in Fig. 4, the isothermal magnetizations under $H\parallel a$ and $H\parallel b$ are linear with increasing field for most temperatures which are consistent with an AFM state. Nevertheless, for $H\parallel c$, nonlinear M(H) curve appears below T$_{m2}$=50~K. Especially for 30~K and 40~K data, they firstly show a ferromagnetic-like fast magnetization to about 0.13$\mu_{B}$/Fe at low field then increases linear with field. This fast magnetization seems to gradually disappear below T$_{m3}$=25~K. The above observations suggest the magnetic transition at T$_{m2}$ is a ferrimagnetic one. The magnetic hysteresis behavior at 35~K in Fig. 5(a) and the peak in the imaginary part of AC susceptibility $\chi''$ at T$_{m2}$ (Fig. 5(b)) further support this conclusion. 

In previous section, the double-$\mathbf{q}$ magnetic structure below T$_{m2}$ identified from neutron diffraction is actually antiferromagnetic-like without net moment component in any direction. This seems to contradict the ferrimagnetic behavior observed in magnetization measurements. The explanation might be, the neutron experiment is carried out in zero field which means that the ground state without magnetic field may actually have no net magnetization. In addition, the net magnetization 0.13$\mu_{B}$/Fe observed in Fig. 3(c) is quite small. This small ferrimagnetic component is not easily detected due to the neutron instrumental resolution. We also need to point out, the moment of Fe has no component along $c$-axis at 80~K but Fe2 actually has a $c$-axis component of 1.14~$\mu_{B}$ at 3.5~K which could be seen from Table S7 and Table S8. This indicates the magnetic anisotropy has notable change below T$_{m2}$.   

Another question that needs to be answered is the nature of magnetic transition at T$_{m3}$=25~K. The experimentally measured specific heat of Fe$_2$SiSe$_4$ is shown in Fig. 6. From the raw data C$_p$(T), the magnetic transitions at T$_{m1}$ and T$_{m2}$ can be identified by the jumps, while there is no detectable feature at T$_{m3}$. Then if using an Einstein model-based curve to account for the phonon part of the specific heat, we can obtain the magnetic specific heat C$_{mag}$ shown by the blue curve in Fig. 6. The C$_{mag}$(T) shows a shoulder at T$_{m3}$=25~K. We found this should-like transition also exists in isostructural Fe$_2$SiO$_4$ at similar temperature (20~K) and its origin has been uncovered as a Schottky anomaly arising from the spin-orbit manifold of the Fe$^{2+}$ ion\cite{Fe2SiO4_2007}. To be specific, for Fe$_2$SiO$_4$, Aronson $et$ $al$. found that the lowest cystal-field splitting t$_{2g}$ level is further split into five states due to spin-orbit coupling. These spin-orbit energy levels are determined through inelastic neutron experiments and have dominate influence on the low temperature physical properties. Then the shoulder in specific heat can be quantitatively calculated and simulated by a Schottky anomaly arising from the spin-orbit excitations\cite{Fe2SiO4_2007}. This explanation may also apply to the anomalous magnetic transition at T$_{m3}$ for Fe$_2$SiSe$_4$. As shown in the inset of Fig. 6, above T$_{m1}$, the magnetic entropy of Fe$_2$SiSe$_4$ is close to the limiting value $2Rln5$ which is calculated by assuming the spin-orbit manifold is fully occupied and there are $2S+1=5$ equally occupied states for each Fe$^{2+}$. This gives further support to the above physical picture.

\begin{figure}
	\includegraphics[width=8cm]{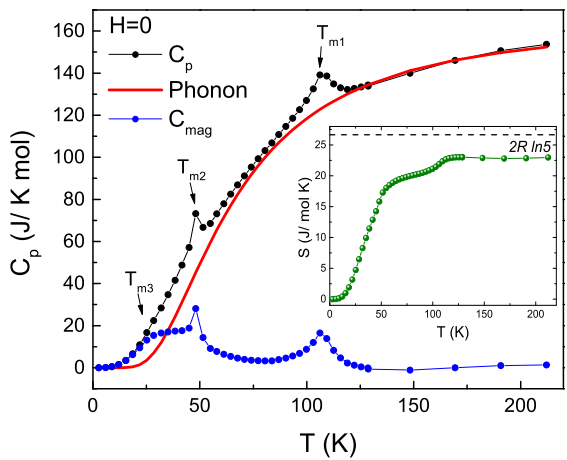}
	\caption {Temperature dependent specific heat of Fe$_2$SiSe$_4$. The magnetic specific heat C$_{mag}$ is derived by subtracting phonon contributions through an Einstein model-based curve fit on the data above 110~K. Then the calculated magnetic entropy is shown in the inset.} \label{Fig6}
\end{figure}

The thermal population in different spin-orbital states may drastically change across T$_{m3}$, which lead to rich magnetic behaviors in Fe$_2$SiSe$_4$. As demonstrated in Fig. 4 and Fig. 5, the magnetization switches from AFM to ferromagnetic behavior then back to AFM behavior as deceasing temperature. In addition, magnetic field induced spin-flop transitions are also observed for M(H) curve at 15~K and 2~K under $H\parallel c$. Combined with the semiconductor nature of this compound which will be discussed in the next section, Fe$_2$SiSe$_4$ may find applications in optoelectronics and magnetic devices.

\subsection{Band structure}

\begin{figure}
	\includegraphics[width=10cm]{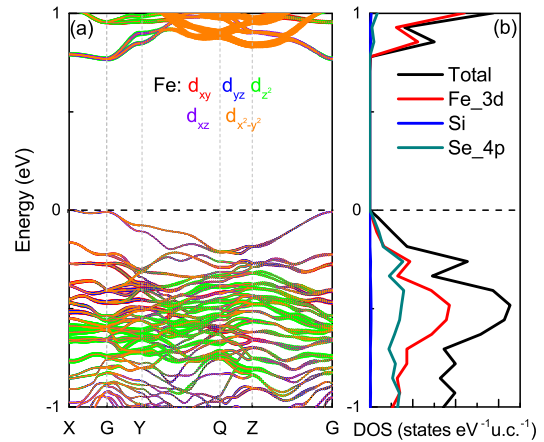}
	\caption {Calculated band structure (a) and total and projected density of states (b) of Fe$_2$SiSe$_4$.} \label{Fig7}
\end{figure}

The sawtooth lattice is known to have flat-bands similar as kagome lattice. The band structure of Fe$_2$SiSe$_4$ obtained by DFT calculations is shown in Fig. 7. The result reveals Fe$_2$SiSe$_4$ is a semiconductor with indirect band gap of 0.7~eV. The gap value is close to that determined from experiment. As shown in section III of Supplemental Material\cite{Supple}, the band gap is determined to be 0.40~eV through fitting the resistance curve and 0.66~eV by the diffusion reflectance spectroscopy (DRS). Considering the errors of different methods, 0.66~eV might be a more accurate value of the band gap of Fe$_2$SiSe$_4$. 

From Fig. 7(a), Fe$_2$SiSe$_4$ is found to have flat bands from the G-X crystallographic direction, similar as other sawtooth material such as Mn$_2$SiSe$_4$ and Fe$_2$GeSe$_4$\cite{Flatband2015,Mn2SiSe4}. The corresponding total and projected density of states (DOS) for Fe$_2$SiSe$_4$ are shown in Fig. 7(b). The states near the Fermi level mainly come from the contributions from Fe 3$d$ states and Se 4$p$ states. Strong hybridization of Se 4$p$ and Fe 3$d$ orbits could be observed below the Fermi level.
 
For kagome metals where the flat-band is quite close to the Fermi level, novel properties including emergent ferromagnetism, anomalous Hall effect, superconductivity and topological phases have been extensively reported in recent years\cite{Fe3Sn2,CoSnS,AHEMn3Ge,CsVSb,Yin2019,Kang2020,Li2021}. We find the band gap of Fe$_2$SiSe$_4$ is not large, there are possibilities that one could use pressure or chemical doping to tune it into a metal. Then it would be worthwhile to check whether a quasi-flat-band near Fermi level would enable realization of other versatile quantum phenomena.

\section{Conclusions}

In summary, we present a comprehensive study on the sawtooth lattice chalcogenide olivines Fe$_2$SiSe$_4$. This material has shown intriguing magnetic properties. Three magnetic transitions are identified as an AFM one at 110~K, a ferrimagnetic one at 50~K and the last one at 25~K possibly due to the spin-orbit excitation. We determined the magnetic structures at different temperatures through neutron diffraction and discover a non-collinear double-$\mathbf{q}$ magnetic order below 50~K. DFT calculations suggest this complex magnetic structure may be due to the competition of AFM interactions on different sides of sawtooth chain combined with the inter-chain exchange interaction. Through band structure calculation and spectral experiment, Fe$_2$SiSe$_4$ is identified as a magnetic semiconductor with indirect band gap of 0.66~eV and quasi-flat-band. We propose that Fe$_2$SiSe$_4$ may provide a new material playground for further researches on magnetic devices and flat-band effect through chemical doping.

\section*{Acknowledgement}
This work was supported by the National Key Research and Development Program of China (No. 2018YFA0306001, 2022YFA1402802), the National Natural Science Foundation of China  (No. 12074426, No. 12004426, No. 11974432, No. 92165204), Shenzhen International Quantum Academy (Grant No. SIQA202102), National Key Scientific Instrument and Equipment Development Project of NSFC (No. 11227906), the Fundamental Research Funds for the Central Universities, and the Research Funds of Renmin University of China (Grants No. 22XNKJ40), NSAF (Grant No. U2030106).

\bibliography{FeSiSe}{}
\end{document}